\documentclass[twocolumn,pra,superscriptaddress,longbibliography]{revtex4-1}
\usepackage{graphicx,amsmath,amssymb}
\usepackage[colorlinks=true, citecolor=blue, urlcolor=blue, linkcolor=red]{hyperref}

\begin{document}
\title{Non-Markovian temperature sensing}

\author{Ze-Zhou Zhang}
\affiliation{ Key Laboratory of Theoretical Physics of Gansu Province, \\
and Lanzhou Center for Theoretical Physics, Lanzhou University, Lanzhou, China}

\author{Wei Wu}
\email{weiwu@lzu.edu.cn}
\affiliation{ Key Laboratory of Theoretical Physics of Gansu Province, \\
and Lanzhou Center for Theoretical Physics, Lanzhou University, Lanzhou, China}

\begin{abstract}
We investigate the sensing performance of a single-qubit quantum thermometer within a non-Markovian dynamical framework. By employing an exactly numerical hierarchical equations of the motion method, we go beyond traditional paradigms of the Born-Markov theory, the pure dephasing mechanism, and the weak-coupling approximation, which were commonly used in many previous studies of quantum thermometry. We find (i) the non-Markovian characteristics may boost the estimation efficiency, (ii) the sensitivity of quantum thermometry can be effectively optimized by engineering the proportions of different coupling operators in the whole sensor-reservoir interaction Hamiltonian, and (iii) a threshold, above which the strong sensor-reservoir coupling can significantly enhance the sensing precision in the long-encoding-time regime. Our results may have certain applications for the high-resolution quantum thermometry.
\end{abstract}

\maketitle

\section{Introduction}\label{sec:sec1}

Temperature is the most fundamental cornerstone in the theory of both classical and quantum thermodynamics. How to measure the temperature of a quantum reservoir with a high sensitivity has recently attracted much attention~\cite{PUGLISI20171,Mehboudi_2019,PhysRevLett.125.080402,PRXQuantum.2.020322,Mok2021,PhysRevResearch.3.013244}. Quantum thermometry pursues a highly precise measuring of temperature, which can surpass the standard bound set by classical statistics, with the help of coherence, quantum squeezing, entanglement, or other quantum resources. In a typical thermometry scheme, a two-level system~\cite{PhysRevA.98.050101,PhysRevA.99.062114,Tamascelli_2020} or a harmonic oscillator~\cite{PhysRevA.96.062103,PhysRevB.98.045101,PhysRevA.103.023303}, working as the quantum sensor, is coupled to the quantum reservoir of interest. Due to the sensor-reservoir interaction, the information about the reservoir's temperature is encoded into the state of the sensor. Then, by measuring a certain sensor's observables, the knowledge about the temperature can be obtained. A highly sensitive quantum thermometer has wide applications in the fields of physics, biology, and material science.

Roughly speaking, there are two totally different quantum thermometers in the present existent temperature sensing schemes: the fully thermalized thermometer~\cite{PRXQuantum.2.020322,PhysRevA.96.062103,PhysRevB.98.045101,PhysRevLett.114.220405,Potts2019fundamentallimits,PhysRevA.102.012204,Latune_2020} and the partly thermalized thermometer~\cite{PhysRevA.98.050101,PhysRevA.99.062114,Tamascelli_2020,PhysRevA.101.032112,PhysRevX.10.011018,PhysRevA.103.012217,PhysRevResearch.3.013244,Kenfack2021,L_pez_V_zquez_2020}. If the sensor-reservoir interaction is weak and the encoding time is sufficiently long, one can regard the sensor as being completely thermalized, namely, the sensor evolves to its thermal equilibrium state which is independent of the encoding time. In many previous treatments~\cite{PRXQuantum.2.020322,PhysRevLett.114.220405,Campbell_2018}, such a thermal equilibrium state is approximately expressed as a canonical Gibbs state at the same temperature as the reservoir. In this case, the temperature can be readout directly from the canonical Gibbs state, and the corresponding temperature uncertainty is related to the heat capacity of the sensor~\cite{PhysRevLett.114.220405,Campbell_2018,Liu_2019}. On the other scenario, if the sensor-reservoir interaction is too strong, resulting in the so-called bound state effect~\cite{Xiong2015,PhysRevA.89.012128,PhysRevE.90.022122}, or the encoding time is short, the sensor cannot be completely thermalized, which implies the state of the sensor is unstable in the time domain. In this situation, to obtain the information of temperature, one needs to monitor the time evolution of the sensor's reduced density matrix. In comparison with the fully thermalized thermometer, the partly thermalized thermometer may have certain advantages, because the quantum coherence can be utilized as the quantum resource to boost the sensing performance, which means the maximum precision is generally achieved out of the thermal equilibrium regime~\cite{PhysRevA.101.032112,Tamascelli_2020,PhysRevA.103.012217}.

However, many previous studies of the partly thermalized thermometer restricted their attentions to the Born-Markov approximation~\cite{PhysRevA.98.050101,PhysRevA.99.062114,L_pez_V_zquez_2020}, the weak-sensor-reservoir-coupling regime~\cite{PhysRevLett.120.140501}, or the pure dephasing encoding mechanism~\cite{PhysRevResearch.3.013244,PhysRevA.101.032112,PhysRevA.103.012217}. Such treatments can provide an analytical result as well as a intuitionistic picture, but inevitably ignore certain important physical phenomena. For example, as reported in Refs.~\cite{PhysRevA.88.035806,PhysRevA.102.032607,PhysRevA.103.L010601,PhysRevApplied.15.054042}, the authors demonstrated that the non-Markovian effect can severely influence the performance of quantum sensing; and the authors of Refs.~\cite{Tamascelli_2020,PhysRevA.102.032607,SALARISEHDARAN2019126006} reported that the pure dephasing mechanism is not the optimal sensor-reservoir encoding scenario. Thus, to obtain a global view and more physical insights into the quantum thermometry, going beyond the above traditional paradigms is highly desirable.

To address these concerns, we employe the hierarchical equations of motion (HEOM) approach~\cite{doi:10.1143/JPSJ.58.101,PhysRevA.41.6676,doi:10.1063/5.0011599,PhysRevE.75.031107,doi:10.1063/1.2713104,doi:10.1063/1.2938087,PhysRevA.98.012110,PhysRevA.98.032116}, which is a non-perturbatively numerical method, to investigate the effect of the reservoir's characteristic, as well as the form of the sensor-reservoir coupling operator on the performance of a single-qubit quantum thermometer in both partly and fully thermalized situations.

This paper is organized as follows. In Sec.~\ref{sec:sec2}, we briefly outline some basic formalism about the quantum parameter estimation. In Sec.~\ref{sec:sec4}, we employe the HEOM method to investigate the sensitivity of a single-qubit quantum thermometry. The conclusion of this paper is drawn in Sec.~\ref{sec:sec5}. In the two appendixes, we provide some additional materials about the HEOM method, the Born-Markov master equation approach, as well as the variational polaron transformation. Throughout the paper, we set $\hbar=k_{\mathrm{B}}=1$ for the sake of simplicity.

\section{Quantum parameter estimation}\label{sec:sec2}

In this section, we would like to recall some basic concepts as well as some important formalism in quantum parameter estimation theory. Generally speaking, to sense a physical quantity $\lambda$ labeling on a quantum system, one first needs a quantum sensor, which is initially prepared in a suitable input state $\varrho_{\mathrm{in}}$, and coupled to the sensor to the system of interest. Due to the sensor-system interaction, the message about $\lambda$ can be encoded into the output state of the sensor via a $\lambda$-dependent mapping $\varrho_{\mathrm{out}}=\mathcal{M}_{\lambda}(\varrho_{\mathrm{in}})=\varrho_{\lambda}$. Such an encoding process can be realized by a unitary rotation~\cite{Hauke2016,PhysRevB.99.045117,PhysRevLett.124.060402} or a nonunitary reduced dynamics~\cite{PhysRevA.102.032607,PhysRevA.103.L010601,PhysRevApplied.15.054042}.

As long as the output state $\varrho_{\lambda}$ is obtained, the information about $\lambda$ can be extracted by measuring the sensor's observable. More specifically speaking, one can construct a measurement operator $\mathcal{\hat{O}}$ and calculate the expected value $\mathcal{\bar{O}}\equiv \mathrm{Tr}(\varrho_{\lambda}\mathcal{\hat{O}})$ as well as the variance $\mathcal{\check{O}}^{2}\equiv \overline{\mathcal{O}^{2}}-\mathcal{\bar{O}}^{2}$. Then, the sensing precision with respect to $\{\varrho_{\lambda},\mathcal{\hat{O}}\}$ can be evaluated via the standard error propagation formula~\cite{PhysRevLett.79.3865,PhysRevLett.123.040402} $\delta^{2}\lambda(\mathcal{\hat{O}})=\mathcal{\check{O}}^{2}/|\partial_{\lambda}\mathcal{\bar{O}}|^{2}$. Running over all the possible measurement schemes, one can find the optimal measurement operator $\mathcal{\hat{O}}_{\mathrm{opt}}$ corresponding to the ultimate sensing precision. The ultimate sensing precision is constrained by the famous quantum Cram$\mathrm{\acute{e}}$r-Rao bound~\cite{PhysRevLett.72.3439,Liu_2019}
\begin{equation}\label{eq:eq1}
\delta^{2}\lambda\geq \delta^{2}\lambda(\mathcal{\hat{O}}_{\mathrm{opt}})=\mathcal{F}_{\lambda}^{-1},
\end{equation}
where $\mathcal{F}_{\lambda}\equiv \mathrm{Tr}(\hat{\varsigma}^{2}\varrho_{\lambda})$ with $\hat{\varsigma}$ determined by $\partial_{\lambda}\varrho_{\lambda}=\frac{1}{2}(\hat{\varsigma}\varrho_{\lambda}+\varrho_{\lambda}\hat{\varsigma})$ is the quantum Fisher information (QFI)~\cite{Liu_2019}. Specifically, if the output state $\varrho_{\lambda}$ is a two-dimensional density matrix described in the Bloch representation, namely, $\varrho_{\lambda}=\frac{1}{2}(\mathbf{1}_{2}+\pmb{r}\cdot\pmb{\hat{\sigma}})$ with $\pmb{r}$ being the Bloch vector and $\pmb{\hat{\sigma}}\equiv(\hat{\sigma}_{x},\hat{\sigma}_{y},\hat{\sigma}_{z})$ being the vector of Pauli matrices, the QFI can be easily calculated via~\cite{PhysRevA.87.022337,Liu_2019}
\begin{equation}\label{eq:eq2}
\mathcal{F}_{\lambda}=|\partial_{\lambda}\pmb{r}|^{2}+\frac{(\pmb{r}\cdot\partial_{\lambda}\pmb{r})^{2}}{1-|\pmb{r}|^{2}}.
\end{equation}
When $\varrho_{\lambda}$ is a pure state, Eq.~(\ref{eq:eq2}) further reduces to $\mathcal{F}_{\lambda}=|\partial_{\lambda}\pmb{r}|^{2}$. Physically speaking, the error propagation formula tells us how to extract the message about $\lambda$ from the output state $\varrho_{\lambda}$ with respect to the given measurement operator $\mathcal{\hat{O}}$, while $\mathcal{F}_{\lambda}$ fully quantifies the most statistical information about $\lambda$ contained in the output state $\varrho_{\lambda}$. There is no general way to find the optimal measurement scheme saturating the best attainable precision determined by the QFI. In this sense, how to construct the optimal measurement scheme is of importance in the research of quantum sensing.

\section{Our scheme}\label{sec:sec4}

\begin{figure}
\centering
\includegraphics[angle=0,width=8cm]{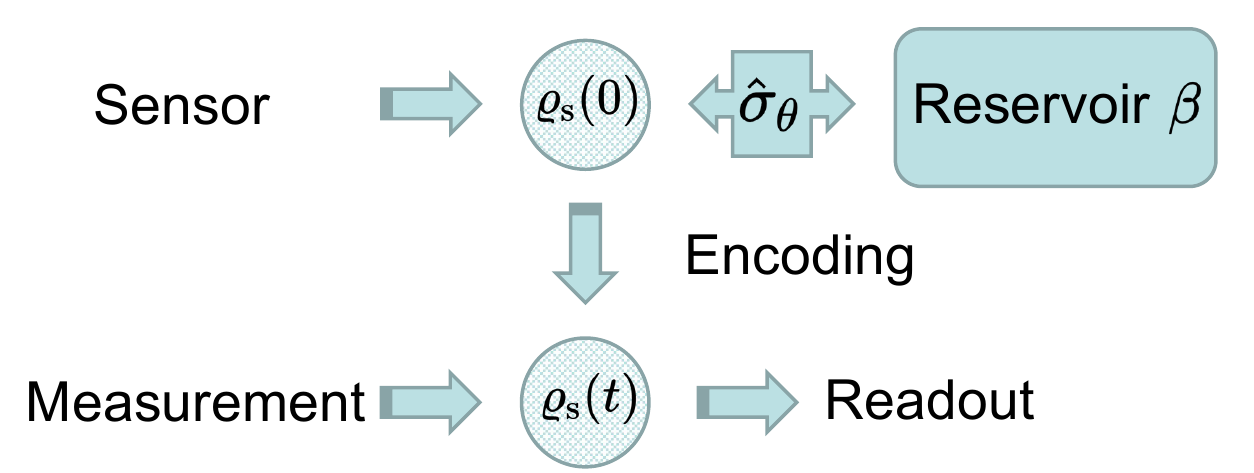}
\caption{Schematic diagram of our quantum thermometry scheme in which a two-level system is used to as the sensor estimate the temperature of a bosonic reservoir.}\label{fig:fig1}
\end{figure}

In this section, we propose our scheme and analyze its sensing efficiency. A two-level system or a qubit, acting as the quantum sensor, is employed to estimate the temperature of a bosonic reservoir (see Fig.~\ref{fig:fig1}). The total Hamiltonian of the sensor plus the finite-temperature reservoir is described as follows:
\begin{equation}\label{eq:eq4}
\hat{H}=\hat{H}_{\mathrm{s}}+\sum_{k}\omega_{k}\hat{b}_{k}^{\dagger}\hat{b}_{k}+\mathcal{\hat{S}}\otimes\sum_{k}g_{k}(\hat{b}_{k}^{\dagger}+\hat{b}_{k}),
\end{equation}
where $\hat{H}_{\mathrm{s}}=\frac{1}{2}\epsilon\hat{\sigma}_{z}+\frac{1}{2}\Delta\hat{\sigma}_{x}$ is the Hamiltonian of the sensor with $\epsilon$ being the bias and $\Delta$ being the tunneling parameter, operators $\hat{b}_{k}$ and $\hat{b}_{k}^{\dagger}$ are annihilation and creation operators of the $k$th bosonic mode with frequency $\omega_{k}$, respectively. The parameter $g_{k}$ labels the coupling strength between the sensor and the $k$th bosonic mode. Generally, it is very convenient to encode the frequency dependence of the interaction strengths in the so-called spectral density, which is defined by $J(\omega)\equiv\sum_{k}g_{k}^{2}\delta(\omega-\omega_{k})$. In this paper, we consider an Ohmic spectral density with a Drude-type cutoff
\begin{equation}\label{eq:eq5}
J(\omega)=\frac{2}{\pi}\frac{\chi\omega\omega_{\mathrm{c}}}{\omega^{2}+\omega_{\mathrm{c}}^{2}},
\end{equation}
where $\chi$ qualifies the sensor-reservoir coupling strength, and $\omega_{\mathrm{c}}$ denotes the cutoff frequency.

In our scheme, we consider a general coupling operator, i.e., $\mathcal{\hat{S}}$ has the form of~\cite{doi:10.1021/acs.jpclett.0c00985}
\begin{equation}\label{eq:eq6}
\mathcal{\hat{S}}=\hat{\sigma}_{\theta}\equiv\sin\theta\hat{\sigma}_{z}+\cos\theta\hat{\sigma}_{x}.
\end{equation}
By varying the coupling angle $\theta$, all the directions on the $x$-$z$ plane of the Bloch sphere can be taken into account. In Refs.~\cite{doi:10.1063/1.4950888,doi:10.1063/1.4825205,doi:10.1063/5.0027976,doi:10.1080/00268976.2018.1430385}, $\sin\theta\hat{\sigma}_{z}$ is named as the diagonal coupling term, while $\cos\theta\hat{\sigma}_{x}$ is called the off-diagonal coupling operator. However, such a naming convention may lead to misunderstanding when the diagonal and the off-diagonal terms are included in both $\hat{H}_{\mathrm{s}}$ and $\mathcal{\hat{S}}$ at the same time. To avoid such a problem, in this paper, we prefer to use the $z$-type and $x$-type coupling terms to indicate $\sin\theta\hat{\sigma}_{z}$ and $\cos\theta\hat{\sigma}_{x}$ in $\mathcal{\hat{S}}$, respectively. Due to the rotational symmetry of $\hat{\sigma}_{\theta}$, we restrict our study to $0\leq\theta<\pi$. If $\theta=\pi/2$, a more $z$-type coupling operator is included in $\mathcal{\hat{S}}$; when $\theta=0$, only the $x$-type coupling is considered~\cite{doi:10.1063/1.4950888}. As demonstrated in Refs.~\cite{doi:10.1063/1.4950888,doi:10.1063/1.3697817,doi:10.1063/1.4825205}, the authors demonstrated the $x$-type coupling term plays a positive role in restraining coherence and can lead to a much richer ground-state phase diagram in the spin-boson systems. These results inspire us to explore the effect of $z$-type and $x$-type coupling on sensitivity of quantum thermometry.

To realize our sensing scheme, we assume the whole sensor-reservoir system is initially prepared in a product state
\begin{equation}\label{eq:eq7}
\begin{split}
\varrho_{\mathrm{sb}}(0)=&\varrho_{\mathrm{s}}(0)\otimes\varrho_{\mathrm{b}}^{\mathrm{G}}\\
=&|\psi_{\mathrm{s}}(0)\rangle\langle\psi_{\mathrm{s}}(0)|\otimes\frac{\exp(-\beta\hat{H}_{\mathrm{b}})}{\mathrm{Tr}_{\mathrm{b}}[\exp(-\beta\hat{H}_{\mathrm{b}})]},
\end{split}
\end{equation}
where $|\psi_{\mathrm{s}}(0)\rangle=\cos\alpha|\mathrm{e}\rangle+\sin\alpha e^{-i\varphi}|g\rangle$ with $|\mathrm{e},g\rangle$ being the eigenstates of the Pauli $z$-operator and $\hat{H}_{\mathrm{b}}=\sum_{k}\omega_{k}\hat{b}_{k}^{\dagger}\hat{b}_{k}$ denotes the Hamiltonian of the reservoir. The parameter $\beta\equiv T^{-1}$ denotes the inverse temperature and is the quantity of interest to be estimated in this paper.

During the purely numerical calculations to the QFI, one needs to handle the first-order derivative to the parameter $\beta$, say $\partial_{\beta}\pmb{r}$ in Eq.~(\ref{eq:eq2}). In this paper, the first-order derivative for an arbitrary $\beta$-dependent function $f_{\beta}$ is numerically done by adopting the following finite difference method:
\begin{equation}\label{eq:eq8}
\partial_{\beta} f_{\beta}\simeq\frac{-f_{\beta+2\delta}+8f_{\beta+\delta}-8f_{\beta-\delta}+f_{\beta-2\delta}}{12\delta}.
\end{equation}
We set $\delta/\beta=10^{-6}$, which provides a very good accuracy.

\subsection{Non-Markovian temperature sensing}\label{sec:sec4a}

In this subsection, we first investigate the dynamical behavior of QFI within an exact non-Markovian framework. Depending on the characteristic of the reservoir spectral density, quantum reservoirs can be roughly classified into two different categories: the fast reservoir $\omega_{\mathrm{c}}>\max\{\epsilon,\Delta\}$ and the slow reservoir $\omega_{\mathrm{c}}\leq\max\{\epsilon,\Delta\}$~\cite{doi:10.1080/00268976.2018.1430385,doi:10.1063/1.4722336,https://doi.org/10.1002/wcms.1375}. An interesting question arises here: what is the influence of the reservoir's characteristic on the sensing performance? In this subsection, we try to address this question.

\begin{figure}
\centering
\includegraphics[angle=0,width=0.35\textwidth]{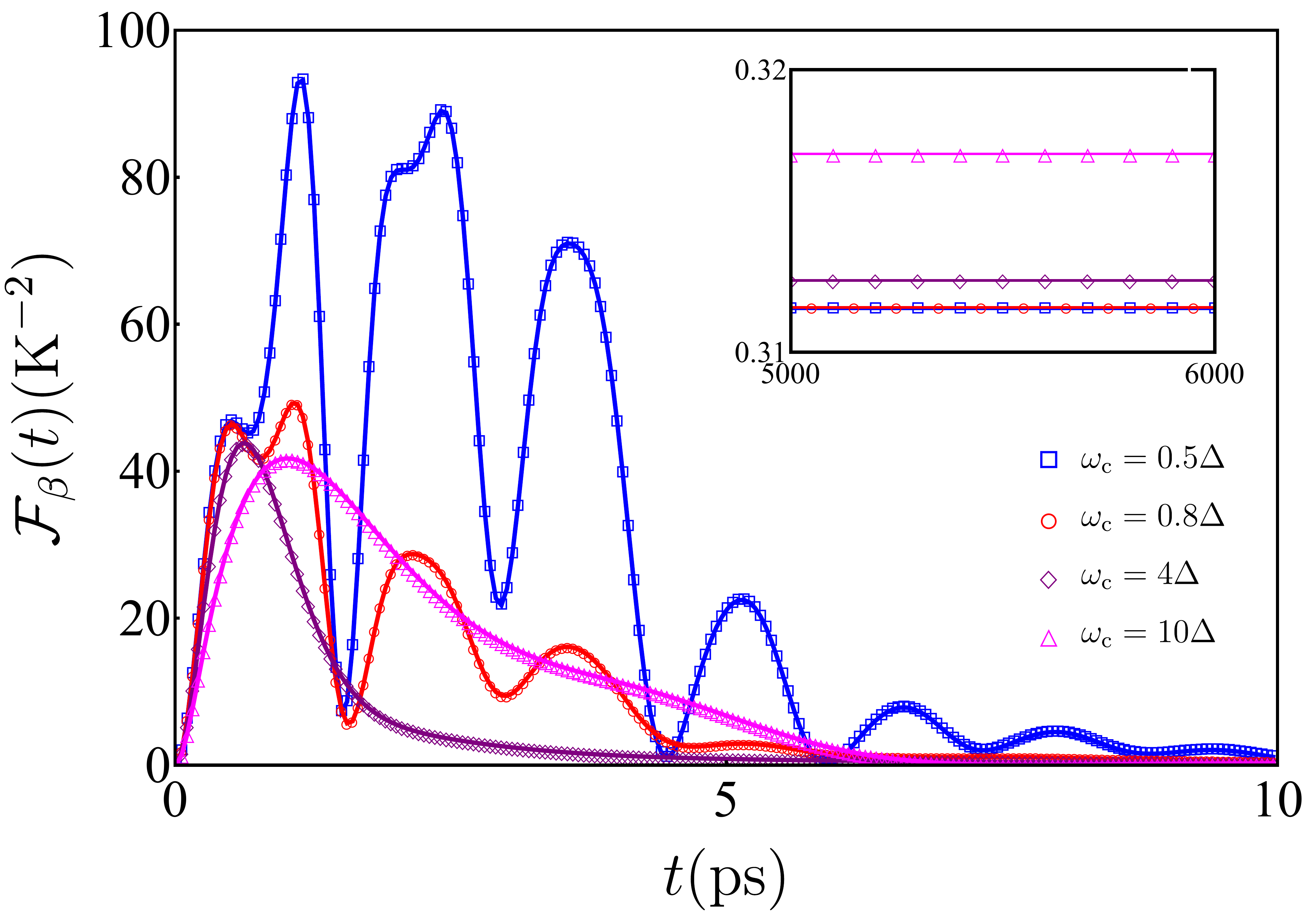}
\caption{The dynamics QFI with different cutoff frequencies: $\omega_{\mathrm{c}}=0.5\Delta$ (blue rectangles), $\omega_{\mathrm{c}}=0.8\Delta$ (red circles), $\omega_{\mathrm{c}}=4\Delta$ (purple rhombuses) and $\omega_{\mathrm{c}}=10\Delta$ (magenta triangles). The inset figure depicts the steady-state QFI in the long-time regime. Parameters are chosen as $\alpha=\pi/4$, $\varphi=\pi/2$, $\epsilon=0.5\Delta$, $\beta\Delta=0.06$, $\theta=0$ and $\chi=0.06\Delta$ with $\Delta=1\,\mathrm{cm}^{-1}$.}\label{fig:fig2}
\end{figure}

\begin{figure}
\centering
\includegraphics[angle=0,width=0.35\textwidth]{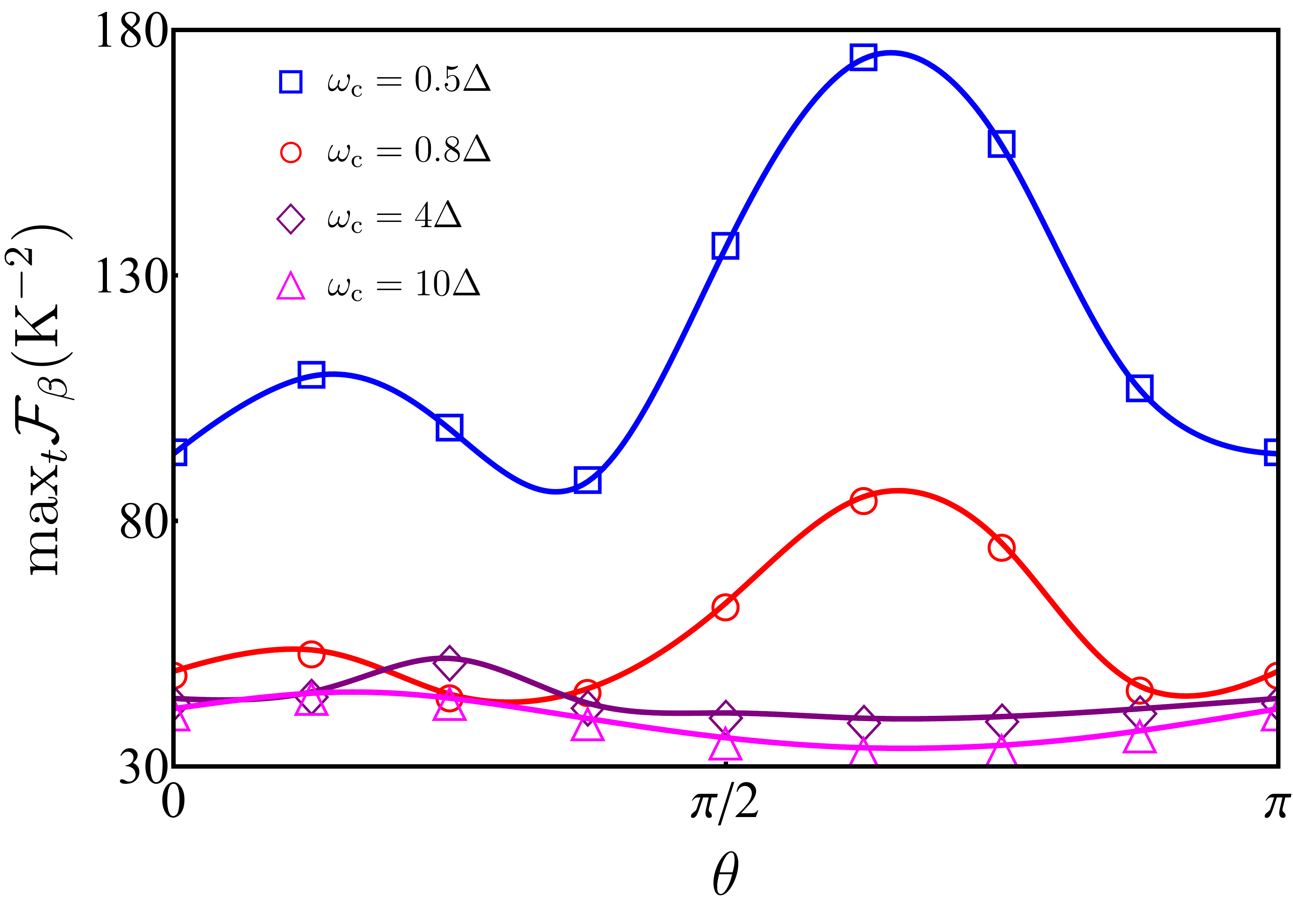}
\caption{The maximum QFI with respect to the optimal encoding time versus the coupling angle for different cutoff frequencies: $\omega_{\mathrm{c}}=0.5\Delta$ (blue rectangles), $\omega_{\mathrm{c}}=0.8\Delta$ (red circles), $\omega_{\mathrm{c}}=4\Delta$ (purple rhombuses) and $\omega_{\mathrm{c}}=10\Delta$ (magenta triangles). Other parameters are the same as those of Fig.~\ref{fig:fig2}.}\label{fig:fig3}
\end{figure}

We first consider the so-called deep fast-reservoir regime, i.e., $\omega_{\mathrm{c}}\gg\max\{\epsilon,\Delta\}$, in which the reservoir is memoryless (see Fig.~\ref{fig:fig6} in Appendix A and Refs.~\cite{doi:10.1080/00268976.2018.1430385,doi:10.1063/1.4722336,https://doi.org/10.1002/wcms.1375,PhysRevA.86.012115} for more details). In this regime, the dynamical behavior of QFI from the HEOM approach is displayed by the magenta triangles in Fig.~\ref{fig:fig2}. One can see the QFI gradually increases from its initial value $\mathcal{F}_{\beta}(0)=0$ to the maximum value. Then, the QFI begins to decrease and eventually tends to a steady value in the long-encoding-time limit. Such behavior of $\mathcal{F}_{\beta}(t)$ is physically reasonable because the whole temperature-sensing process includes the following three steps. First, no message about the temperature is contained in the initial state of the sensor, resulting in $\mathcal{F}_{\beta}(0)=0$ at the beginning. Then, the sensor-reservoir interaction, which generates the temperature's information in $\varrho_{\mathrm{s}}(t)$ with the evolution of encoding time, leads to the increase of QFI. Finally, in the long-time limit, the value of $\mathcal{F}_{\beta}(t)$ remains unchanged because, in this step, the message about the temperature is completely from the steady-state reduced density matrix $\varrho_{\mathrm{s}}(\infty)$, which is independent of the encoding time. The above result means there exists an optimal encoding time which can maximize the value of QFI. The appearance of maximal QFI can be understood as the physical result of the interplay between the encoding and the decoherence ~\cite{PhysRevA.102.032607,PhysRevA.103.L010601,PhysRevApplied.15.054042,PhysRevA.97.012125}. In Fig.~\ref{fig:fig3}, we plot the maximal QFI, $\max_{t}\mathcal{F}_{\beta}$, as a function of the coupling angle $\theta$. One can see the maximal QFI is quite sensitive to the coupling angle. Using our above results, one can design the most efficient sensor-reservoir interaction Hamiltonian to obtain the maximum precision estimation by adjusting the weight of the $x$-type coupling operator in the whole sensor-reservoir interaction Hamiltonian.

Next, we consider the slow-reservoir regime, for example, the blue rectangles with $\omega_{\mathrm{c}}=0.5\Delta$ in Fig.~\ref{fig:fig2}, in which the reservoir-induced decoherence should be strongly non-Markovian~\cite{doi:10.1080/00268976.2018.1430385,doi:10.1063/1.4722336,https://doi.org/10.1002/wcms.1375}. In this situation, we find the dynamics of QFI presented by the HEOM method exhibits a collapse-and-revival phenomenon resulting in multiple local maximums, which is completely different from that of the fast reservoir case. A similar result was also reported in Refs.~\cite{PhysRevA.88.035806,PhysRevA.102.032607} and can be regarded as evidence of the information's backflow from the reservoir back to the sensor. Moreover, we also display the maximum QFI with respect to the optimal encoding time versus the coupling angle in Fig.~\ref{fig:fig3}. Multiple local-most-efficient coupling angles are revealed (see the the blue rectangles in Fig.~\ref{fig:fig3}). Additionally, we find the maximal QFI in the slow-reservoir regime can be much larger than that of the fast-reservoir case for the entire range of $0\leq\theta<\pi$. This result suggests the non-Markovian effect, generated by the characteristics of slow reservoir, may be used to improve the estimation precision.

\subsection{Breakdown of the canonical statistics}\label{sec:sec4b}

In this subsection, we shall discuss the feature of steady-state QFI in the long-encoding-time regime. If the sensor-reservoir coupling is weak, the long-time steady state of a fully thermalized thermometer can be described by the canonical Gibbs state, namely, $\varrho_{\mathrm{s}}(\infty)\simeq\varrho_{\mathrm{s}}^{\mathrm{G}}=\exp(-\beta \hat{H}_{\mathrm{s}})/\mathrm{Tr}[\exp(-\beta \hat{H}_{\mathrm{s}})]$. In such an approximate treatment, by diagonalizing $\hat{H}_{\mathrm{s}}$, one can easily find the ratio of $\varrho_{\mathrm{gg}}(\infty)/\varrho_{\mathrm{ee}}(\infty)$ is a monotonously and exponentially increasing function
\begin{equation}\label{eq:eq9}
\frac{\varrho_{\mathrm{gg}}(\infty)}{\varrho_{\mathrm{ee}}(\infty)}\simeq\frac{\varrho_{\mathrm{gg}}^{\mathrm{G}}}{\varrho_{\mathrm{ee}}^{\mathrm{G}}}=e^{\beta\Omega},
\end{equation}
where $\Omega=\sqrt{\epsilon^{2}+\Delta^{2}}$ shall be regarded as the sensor's Rabi frequency. And the corresponding QFI with respect to the canonical Gibbs state is given by
\begin{equation}\label{eq:eq10}
\mathcal{F}_{\beta}^{\mathrm{G}}=\frac{\Omega^{2}}{2+2\cosh(\beta\Omega)}.
\end{equation}
From the above expression, one can easily find $\mathcal{F}_{\beta}^{\mathrm{G}}(\Omega)$ is not a monotonic function. With a fixed temperature, $\mathcal{F}_{\beta}^{\mathrm{G}}(\Omega)$ is a increasing function for $\Omega\in[0,\Omega^{*})$ with $\Omega^{*}$ being determined by $\partial_{\Omega}\mathcal{F}_{\beta}^{\mathrm{G}}(\Omega)=0$; while, for $\Omega\in[\Omega^{*},\infty)$, $\mathcal{F}_{\beta}^{\mathrm{G}}(\Omega)$ monotonically decreases. The above analysis means the steady-state behaviors of the population ratio and the QFI strongly depend on the sensor's Rabi frequency $\Omega$.

However, as demonstrated in many previous articles~\cite{doi:10.1063/1.4722336,PhysRevE.80.041106,PhysRevE.84.011114,PhysRevE.84.051112}, in the deep slow-reservoir regime, which generally induces a strongly non-Markovian effect, the reservoir-induced decoherence can give rise to a drastic frequency renormalization, namely, $\Omega\rightarrow\tilde{\Omega}$. The renormalized frequency $\tilde{\Omega}$ is smaller than the original frequency $\Omega$ and their deviation becomes larger as the increase of the sensor-reservoir coupling~\cite{doi:10.1063/1.4722336,PhysRevE.80.041106,PhysRevE.84.011114,PhysRevE.84.051112,doi:10.1063/1.447055}. This result can be qualitatively understood by performing a polaron transformation. Making use of the polaron transformation, as displayed in Appendix B, the sensor's Rabi frequency is renormalized as
\begin{equation}\label{eq:eq12}
\tilde{\Omega}_{\mathrm{P}}=\sqrt{\tilde{\epsilon}^{2}+\eta^{2}\tilde{\Delta}^{2}},
\end{equation}
where $\tilde{\epsilon}=\epsilon\sin\theta+\Delta\cos\theta$, $\tilde{\Delta}=\Delta\sin\theta-\epsilon\cos\theta$, and $\eta$ is the renormalized factor, whose explicit expression is given by Eq.~(\ref{eq:eqb8}), induced by the polaron transformation. In Table~\ref{table:table1}, we display the frequency renormalization within the framework of polaron transformation theory $\tilde{\Omega}_{\mathrm{P}}/\Omega$ versus the sensor-reservoir coupling strength $\chi$. It is clear to see a larger $\chi$ leads to a smaller $\tilde{\Omega}_{\mathrm{P}}/\Omega$. Such a frequency renormalization phenomenon is responsible for the so-called localized-delocalized transition in the well-known spin-boson model~\cite{PhysRevE.80.041106,PhysRevE.84.011114}.

\begin{table}[t]
\begin{center}
\caption{The frequency renormalization $\tilde{\Omega}_{\mathrm{P}}$ and $\tilde{\Omega}_{\mathrm{H}}$ versus the sensor-reservoir coupling strength $\chi$. Parameters are chosen as $\alpha=\pi/4$, $\varphi=\pi/2$, $\theta=2\pi/3$, $\epsilon=2\Delta$, $\beta\Delta=0.95$, $\omega_{\mathrm{c}}=0.8\Delta$, and $\Delta=0.1\mathrm{cm}^{-1}$.}\label{table:table1}
\setlength{\tabcolsep}{7.25pt}
\begin{tabular}{ccccccc}
\hline
  \hline
  $\chi/\Delta$ & 0 & 0.1 & 0.2 & 0.3 & 0.4 & 0.5\\
  $\tilde{\Omega}_{\mathrm{P}}/\Omega$ & 1.000 & 0.968 & 0.932 & 0.890 & 0.837 & 0.741\\
  $\tilde{\Omega}_{\mathrm{H}}/\Omega$ & 1.000 & 0.834 & 0.818 & 0.794 & 0.766 & 0.734\\
  \hline
  \hline
\end{tabular}
\end{center}
\end{table}

\begin{figure}
\centering
\includegraphics[angle=0,width=0.49\textwidth]{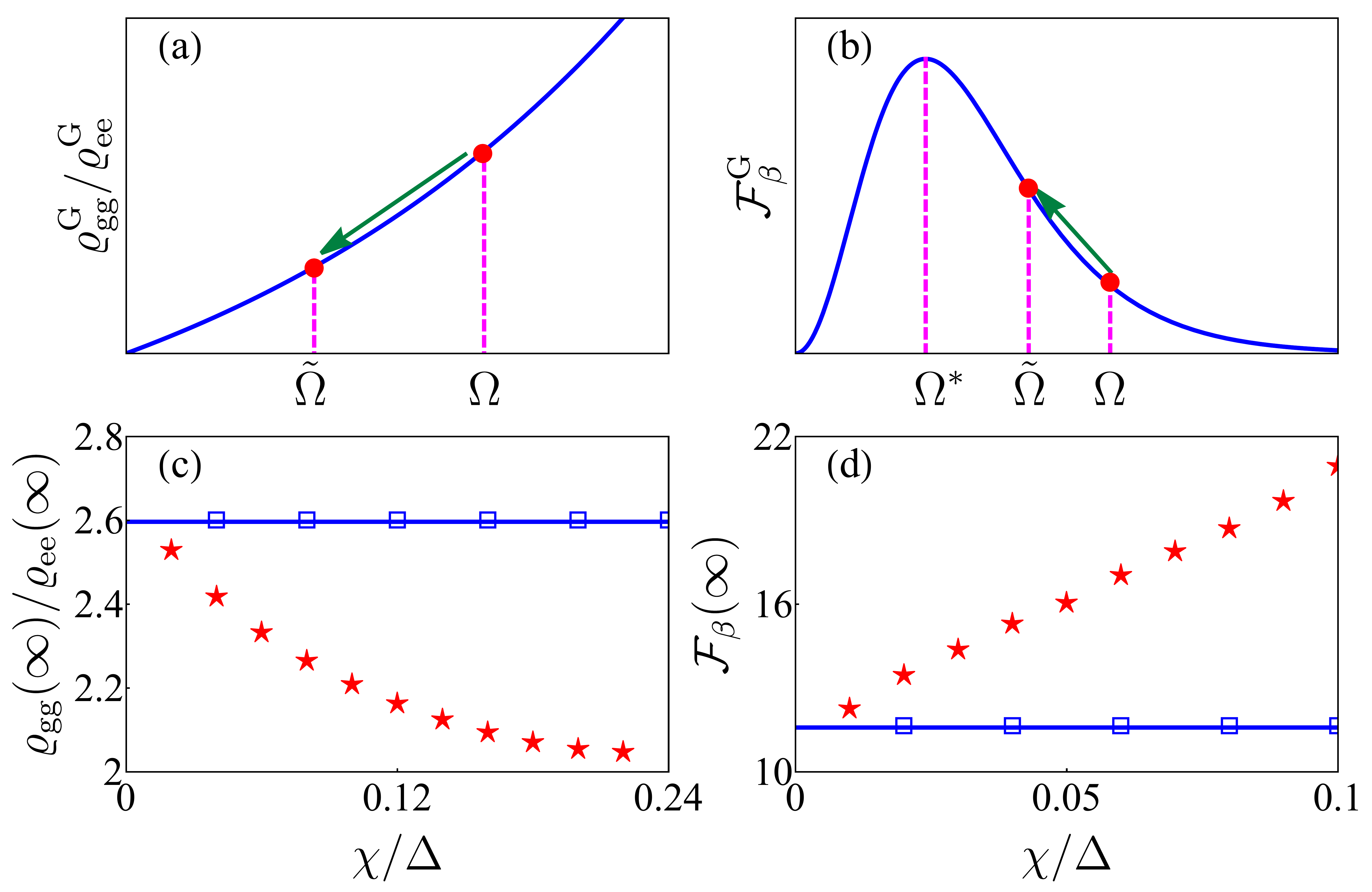}
\caption{Schematic diagram of the influence of frequency renormalization $\Omega\rightarrow\tilde{\Omega}$ on (a) the steady-state population ratio and (b) the QFI. (c) The steady-state population ratio $\varrho_{\mathrm{gg}}(\infty)/\varrho_{\mathrm{ee}}(\infty)$ is plotted as a function of the sensor-reservoir coupling strength $\chi$. (d) The steady-state QFI versus $\chi$. The red five-point stars are the numerical results obtained by the HEOM method, the blue rectangles are the steady-state solutions from the Born-Markov master equation approach, while the blue solid lines represent the results from the canonical Gibbs state. Parameters are chosen as $\alpha=\pi/4$, $\varphi=\pi/2$, $\epsilon=0.5\Delta$, $\beta\Delta=5$, and $\omega_{\mathrm{c}}=0.5\Delta$ with $\Delta=50\,\mathrm{cm}^{-1}$. Here, $\Omega^{*}\simeq24\,\mathrm{cm}^{-1}$ and $\Omega\simeq 56\,\mathrm{cm}^{-1}$.}\label{fig:fig4}
\end{figure}

\begin{figure}
\centering
\includegraphics[angle=0,width=0.49\textwidth]{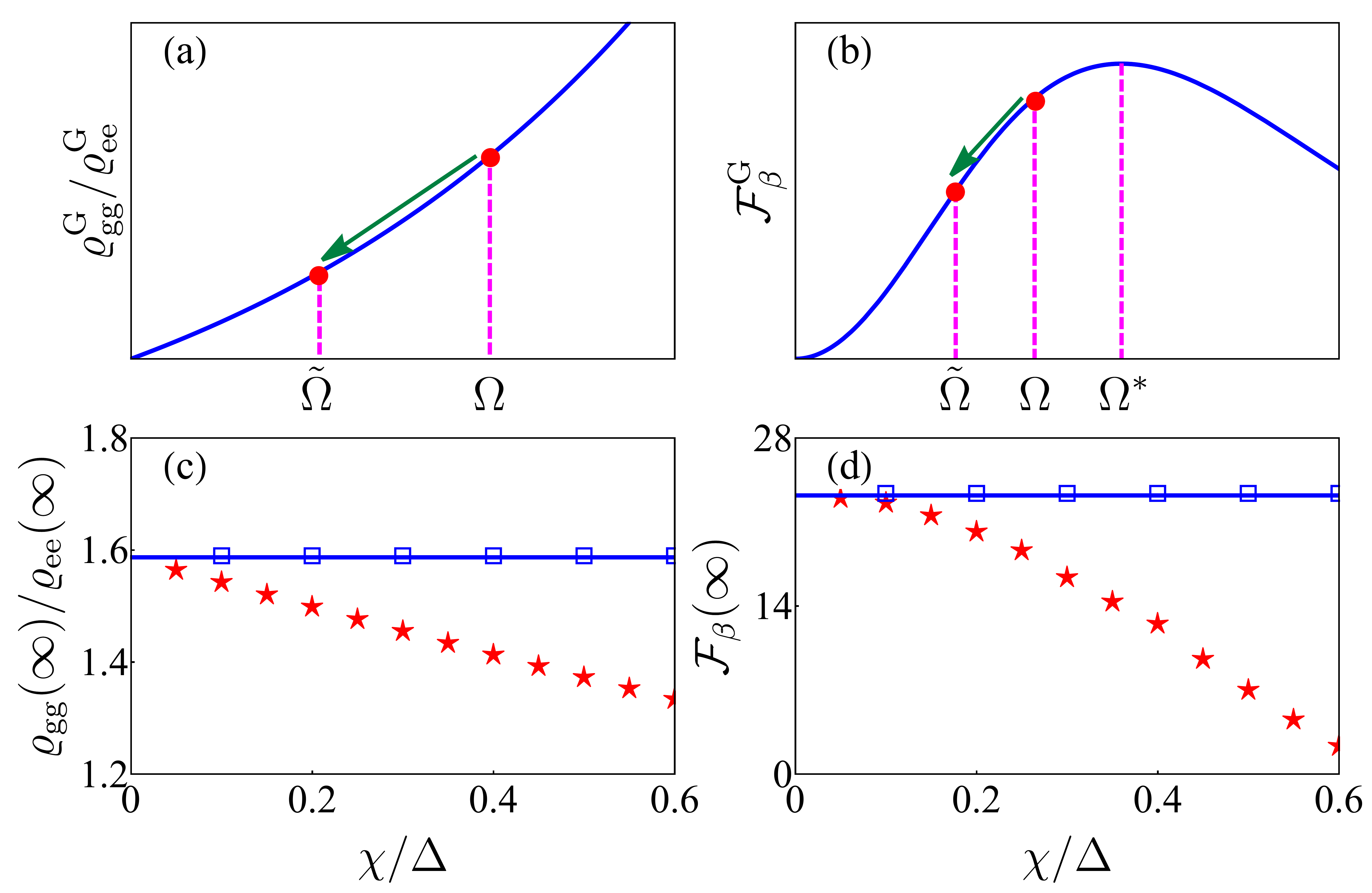}
\caption{The same with Fig.~\ref{fig:fig4}, but $\beta\Delta=1$ and $\Delta=10\,\mathrm{cm}^{-1}$. Here, $\Omega^{*}\simeq24\,\mathrm{cm}^{-1}$ and $\Omega\simeq 11\,\mathrm{cm}^{-1}$.}\label{fig:fig5}
\end{figure}

Thus, the most simple and straightforward correction to the previous assumption of canonical statistics is replacing the original Rabi frequency $\Omega$ by a smaller renormalized frequency $\tilde{\Omega}$. Under such treatment, one can see the frequency shift $\Omega\rightarrow\tilde{\Omega}$ shall lead to a smaller value of $\varrho_{\mathrm{gg}}(\infty)/\varrho_{\mathrm{ee}}(\infty)$ [see Figs.~\ref{fig:fig4}(a) and 5(a)] compared with that of the canonical Gibbs state case, which implies the breakdown of canonical statistics~\cite{Zhu_2020,PhysRevA.90.032114,doi:10.1063/1.4940218}. Moreover, if $\Omega>\Omega^{*}$, the frequency renormalization results in a larger QFI in comparison with that of the canonical statistics [see Fig.~\ref{fig:fig4}(b)]. On the contrary, if $\Omega<\Omega^{*}$, the frequency renormalization in turn diminishes the sensing precision [see Fig.~\ref{fig:fig5}(b)].

The above analytical analysis is completely based on the polaron transformation, which only provides a qualitative interpretation as well as a pictorial illustration. To obtain a quantitative result, we display the steady-state population ratio and the QFI as functions of the coupling strength from the HEOM and the Born-Markov master equation methods (see Appendix A for details) in Figs.~\ref{fig:fig4}(c,) and 4(d) and Figs.~\ref{fig:fig5}(c,) and 5(d). As a result of neglecting the higher-order terms of the sensor-reservior interaction, the steady-state solutions from the Born-Markov master equation approach coincide with the canonical statistics, i.e., the steady-state population ratio and the QFI are independent of the coupling strength. In sharp contrast to this, one can see the numerical results obtained by the HEOM method are in good agreement with our previous discussions about the frequency renormalization within the framework of the polaron transformation. In Table~\ref{table:table1}, we also display the renormalized Rabi frequency predicted by the HEOM method, namely $\tilde{\Omega}_{\mathrm{H}}\equiv\beta^{-1}\ln[\varrho_{\mathrm{gg}}(\infty)/\varrho_{\mathrm{ee}}(\infty)]$, as the function of the coupling strength $\chi$. It is clear to see the evolutionary trend of $\tilde{\Omega}_{\mathrm{H}}/\Omega$ in qualitative agreement with that of $\tilde{\Omega}_{\mathrm{P}}/\Omega$. These results demonstrate, in the strong-coupling regime, the noncanonical distribution occurs, which plays a complicated role in the quantum thermometry. The result presented by the red stars in Figs.~\ref{fig:fig4}(c,) and 4(d) and Figs.~\ref{fig:fig5}(c,)and 5(d) cannot be predicted by using the Gorini-Kossakowski-Sudarshan-Lindblad master equation formula or other common perturbative methods, where the effect of frequency shift is generally washed out~\cite{HUANG2010256,PhysRevA.94.062116}.

Therefore, we uncover a threshold $\Omega=\Omega^{*}$, above such critical Rabi frequency, i.e., $\Omega>\Omega^{*}$, the sensing precision can be effectively enhanced by strengthening the sensor-reservoir coupling. In this sense, the sensitivity of quantum thermometry can be improved by engineering the sensor's bare frequency. Here, we concentrate the frequency renormalization mechanism on the slow-reservoir regime, the reason is twofold. First, in the deep fast-reservoir regime, the decoherence is Markovian which means the canonical statistics assumption is valid with no need for corrections. Second, as $\omega_{\mathrm{c}}/\max\{\epsilon,\Delta\}\rightarrow\infty$, the renormalized factor $\eta$ approaches to $1$ accordingly, resulting in the disappearance of frequency shift phenomenon in the deep-reservoir regime. Our result is consistent with that of Ref.~\cite{PhysRevA.96.062103}, in which the authors found the sensing precision of a Brownian thermometer can be improved by increasing the coupling at low temperature. Their result can be easily understood by using our threshold mechanism: the critical Rabi frequency $\Omega^{*}$ approaches to zero in the low-temperature limit, which means the condition $\Omega>\Omega^{*}$ becomes very easy to be satisfied. Thus, as long as the Brownian sensor has a finite bare frequency, i.e., $\Omega>0$, the sensor-reservoir can boost the thermometric precision without doubt. Of course, it is worth mentioning that our scenario is totally different from that found in Ref.~\cite{PhysRevA.96.062103}. Their quantum thermometry is based on the Caldeira-Leggett Hamiltonian, namely, the sensor in Ref.~\cite{PhysRevA.96.062103} is a continuous-variable system, rather than the discrete-variable sensor (qubit) in this paper.

\section{Conclusion}\label{sec:sec5}

In summary, going beyond the usual assumptions of the Born-Markov theory, the pure dephasing mechanism and the weak-coupling approximation, we propose a single-qubit thermometer scheme and analyze its sensing performance with the help of the HEOM approach. We find, when the encoding time is short and the sensor is partly thermalized, the non-Markovian effect induced by the slow reservoir's characteristics may enhance the efficiency of the quantum thermometry. By optimizing the encoding time, we investigate the relation between the maximal QFI and the sensor-reservoir coupling angle. It is revealed that the performance of our quantum thermometry can be further boosted by modulating the weight of the $x$-type coupling operator in the whole sensor-reservoir Hamiltonian. Moreover, in the slow-reservoir regime, by strengthening the sensor-reservoir coupling, we find the noncanonical feature appears in the sensor's steady state and the corresponding QFI obtained by the HEOM method can be larger than the value predicated by the canonical Gibbs state as well as the Born-Markov master equation approach, under suitable conditions. A possible threshold mechanism, which is based on the frequency renormalization appearing in the strongly non-Markovian regime, is proposed to explain the above result in the long-encoding-time limit. Finally, due to the generality of the qubit-based quantum thermometer, we expect our results to be of interest for certain potential applications in the researches of quantum metrology and quantum sensing.

\section{Acknowledgments}

The authors thank Dr. Si-Yuan Bai, Professor Jun-Hong An and Professor Hong-Gang Luo for many fruitful discussions. W. Wu wishes to especially thank Dr. Chong Chen and an anonymous referee for their suggestions on the discussion of frequency renormalization. The work was supported by the National Natural Science Foundation (Grant No. 11704025).

\section{Appendix A: HEOM}

The HEOM method is a purely numerical technique, which exactly maps the Schr$\ddot{\mathrm{o}}$dinger equation or quantum Liouville equation to a set of ordinary differential equations by introducing auxiliary density matrices. These ordinary differential equations can be treated numerically by the Runge-Kutta algorithm. Without invoking the Born-Markov, weak-coupling and rotating-wave approximations, the HEOM can provide a rigorous numerical dynamics for an open quantum system~\cite{doi:10.1143/JPSJ.58.101,PhysRevA.41.6676,doi:10.1063/5.0011599,PhysRevE.75.031107,doi:10.1063/1.2713104,doi:10.1063/1.2938087,PhysRevA.98.012110,PhysRevA.98.032116}.

To realize the above HEOM method, it is generally required that the reservoir correlation function $C(t)$, which is defined by
\setcounter{equation}{0}
\renewcommand\theequation{A\arabic{equation}}
\begin{equation}\label{eq:eqa1}
\begin{split}
C(t)=&\mathrm{Tr}_{\mathrm{b}}\Big{(}e^{it\hat{H}_{\mathrm{b}}}\mathcal{\hat{B}}e^{-it\hat{H}_{\mathrm{b}}}\mathcal{\hat{B}}\Big{)}\\
=&\int_{0}^{\infty} d\omega J(\omega)\Big{[}\coth\Big{(}\frac{\beta\omega}{2}\Big{)}\cos(\omega t)-i\sin(\omega t)\Big{]},
\end{split}
\end{equation}
can be expressed as a sum of exponential functions~\cite{doi:10.1143/JPSJ.58.101,PhysRevA.41.6676,doi:10.1063/5.0011599,PhysRevE.75.031107,doi:10.1063/1.2713104,doi:10.1063/1.2938087,PhysRevA.98.012110,PhysRevA.98.032116,doi:10.1063/1.4893931}. Here, $\mathcal{\hat{B}}\equiv\sum_{k}g_{k}(\hat{b}_{k}^{\dagger}+\hat{b}_{k})$. Fortunately, for the Ohmic spectral density considered in this paper, the reservoir correlation function satisfies such requirement. Substituting Eq.~(\ref{eq:eq5}) into Eq.~(\ref{eq:eqa1}), one can find
\begin{equation}\label{eq:eqa2}
C(t)=\sum_{\ell=0}^{\varepsilon}\zeta_{\ell}e^{-\upsilon_{\ell}t},
\end{equation}
where $\upsilon_{\ell}=\omega_{\mathrm{c}}\delta_{0\ell}+2\ell\pi(1-\delta_{0\ell})\beta^{-1}$ denotes the $\ell$th Matsubara frequency and
\begin{equation}
\begin{split}
\zeta_{\ell}=&\frac{4\chi\omega_{\mathrm{c}}}{\beta}\frac{\upsilon_{\ell}}{\upsilon_{\ell}^{2}-\omega_{\mathrm{c}}^{2}}(1-\delta_{0\ell})\\
&+\Big{[}\chi\omega_{\mathrm{c}}\cot\Big{(}\frac{\beta\omega_{\mathrm{c}}}{2}\Big{)}-i\chi\omega_{\mathrm{c}}\Big{]}\delta_{0\ell},
\end{split}
\end{equation}
are the expansion coefficients. Here, we set a truncation parameter $\varepsilon$ to ensure the above series remain finite. With the help of the Eq.~(\ref{eq:eqa2}), the hierarchical equations can be constructed following detailed exposition in Refs.~\cite{PhysRevA.98.012110,PhysRevA.98.032116}. The HEOM reads
\begin{equation}
\begin{split}
\frac{d}{dt}\varrho_{\vec{\nu}}(t)=&\Big{(}\mathcal{\hat{L}}_{\mathrm{s}}-\vec{\nu}\cdot\vec{\mu}\Big{)}\varrho_{\vec{\nu}}(t)\\
&+\hat{\Phi}\sum_{\ell=0}^{\varepsilon}\hat{\varrho}_{\vec{\nu}+\vec{e}_{\ell}}(t)+\sum_{\ell=0}^{\varepsilon}\nu_{\ell}\hat{\Theta}_{\ell}\hat{\varrho}_{\vec{\nu}-\vec{e}_{\ell}}(t),
\end{split}
\end{equation}
where $\vec{\nu}=(\nu_{0},\nu_{1},\nu_{2},.~.~ .,\nu_{\varepsilon})$ is a $(\varepsilon+1)$-dimensional index, $\vec{e}_{\ell}=(0,0,0,.~.~.1_{\ell},.~.~.,0)$ and $\vec{\mu}=(\upsilon_{0},\upsilon_{1},\upsilon_{2},.~.~.,\upsilon_{\varepsilon})$ are $(\varepsilon+1)$-dimensional vectors. The superoperators $\mathcal{\hat{L}}_{\mathrm{s}}$, $\hat{\Phi}$, and $\hat{\Theta}_{\ell}$ are defined as $\mathcal{\hat{L}}_{\mathrm{s}}\equiv-i\hat{H}_{\mathrm{s}}^{\times}$, $\hat{\Phi}\equiv-i\mathcal{\hat{S}}^{\times}$ and
\begin{equation}
\hat{\Theta}_{\ell}=-i\Big{[}\mathrm{Re}(\zeta_{\ell})\mathcal{\hat{S}}^{\times}+i\mathrm{Im}(\zeta_{\ell})\mathcal{\hat{S}}^{\circ}\Big{]},
\end{equation}
where $\hat{o}_{1}^{\times}\hat{o}_{2}\equiv\hat{o}_{1}\hat{o}_{2}-\hat{o}_{2}\hat{o}_{1}$ and $\hat{o}_{1}^{\circ}\hat{o}_{2}\equiv\hat{o}_{1}\hat{o}_{2}+\hat{o}_{2}\hat{o}_{1}$.

In numerical simulations, the initial states of the auxiliary operators are given by
\begin{equation}
\varrho_{\vec{\nu}=\vec{0}}(0)=\varrho_{\mathrm{s}}(0);~~\varrho_{\vec{\nu}\neq\vec{0}}(0)=\vec{0},
\end{equation}
where $\vec{0}=(0,0,0. . .,0)$ is a $(\varepsilon+1)$-dimensional zero vector. In this paper, we constantly increasing the number of the differential equations as well as the value of $\varepsilon$ until the final result converges.

\begin{figure*}
\centering
\includegraphics[angle=0,width=0.99\textwidth]{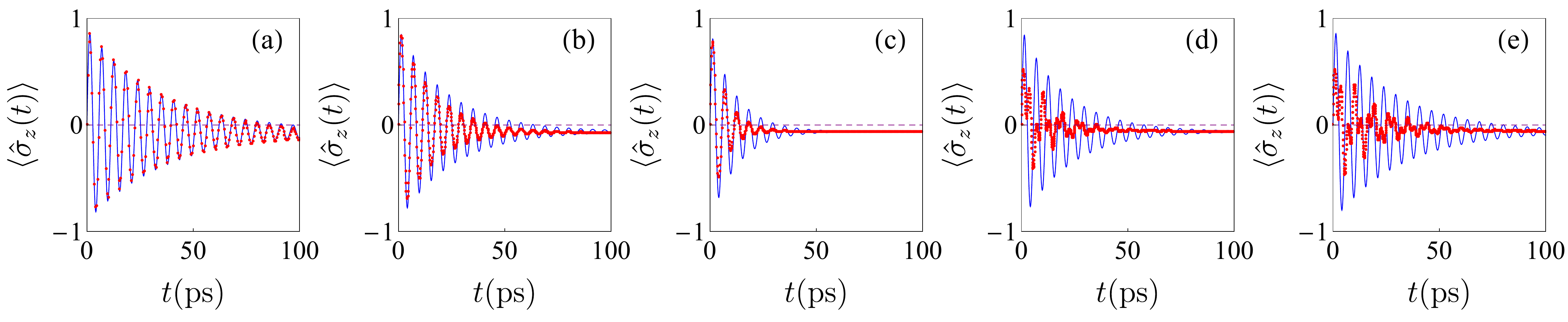}
\caption{The dynamics of population difference $\langle\hat{\sigma}_{z}(t)\rangle$ with different cutoff frequencies: (a) $\omega_{\mathrm{c}}=30\Delta$ (the deep fast-reservoir regime), (b) $\omega_{\mathrm{c}}=20\Delta$, (c) $\omega_{\mathrm{c}}=10\Delta$, (d) $\omega_{\mathrm{c}}=0.07\Delta$, and (e) $\omega_{\mathrm{c}}=0.05\Delta$ (the deep slow-reservoir regime). The red circles are numerical results predicted by the HEOM method, while the blue solid lines are from the Born-Markovian master equation approach. Other parameters are chosen as $\alpha=\pi/4$, $\varphi=\pi/2$, $\epsilon=0.5\Delta$, $\chi=0.06\Delta$, $\beta\Delta=0.25$, $\theta=3\pi/8$, and $\Delta=1\,\mathrm{cm}^{-1}$.}\label{fig:fig6}
\end{figure*}

As a benchmark, we also employe the most common Born-Markov master equation to compute the reduced dynamics of the sensor. As displayed in Ref.~\cite{PhysRevA.94.062116}, the Born-Markov master equation reads
\begin{equation}\label{eq:eqa6}
\frac{d}{dt}\varrho_{\mathrm{s}}(t)=\Big{(}\mathcal{\hat{L}}_{\mathrm{s}}-\mathcal{\hat{S}}^{\times}\hat{\Upsilon}^{\times}+\mathcal{\hat{S}}^{\times}\hat{\Xi}^{\circ}\Big{)}\varrho_{\mathrm{s}}(t),
\end{equation}
where
\begin{equation*}
\hat{\Upsilon}\equiv\int_{0}^{\infty}dt C_{\mathrm{R}}(t)\mathcal{\hat{S}}(-t),
\end{equation*}
and
\begin{equation*}
\hat{\Xi}\equiv-i\int_{0}^{\infty}dt C_{\mathrm{I}}(t)\mathcal{\hat{S}}(-t).
\end{equation*}
Here, $\mathcal{\hat{S}}(t)\equiv e^{it\hat{H}_{\mathrm{s}}}\hat{\mathcal{S}}e^{-it\hat{H}_{\mathrm{s}}}$ and $C_{\mathrm{R}(\mathrm{I})}(t)$ denotes the real (imaginary) part of the correlation function. By neglecting the imaginary Lamb-shift terms~\cite{PhysRevA.94.062116}, the master equation given by Eq.~(\ref{eq:eqa6}) can be further simplified, which shall be very convenient for numerical simulations in practice.

To verify the feasibility of the HEOM method, we here make a comparison between the purely numerical result obtained by the numerical HEOM method and that of the Born-Markov master equation approach. In Fig.~\ref{fig:fig6}, we display the evolution of population difference
\begin{equation}
\langle\hat{\sigma}_{z}(t)\rangle\equiv\mathrm{Tr}\Big{[}\hat{\sigma}_{z}e^{-i\hat{H}t}\varrho_{\mathrm{s}}(0)\otimes\varrho_{\mathrm{b}}^{\mathrm{G}}e^{i\hat{H}t}\Big{]}.
\end{equation}
Good agreement is found between results from the two different approaches in the deep fast-reservoir regime, see Fig.~\ref{fig:fig6}(a). On the contrary, a relatively large deviation is found in the deep slow-reservoir regime, see Figs.~\ref{fig:fig6}6(d) and 6(e). Moreover, we observe that such a deviation becomes more and more evident as $\omega_{\mathrm{c}}/\Delta$ approaches to zero with fixed $\epsilon=0.5\Delta$. This result is physically reasonable, because the reservoir is memoryless in the deep fast-reservoir regime. While, with the decrease of $\omega_{\mathrm{c}}/\Delta$, the non-Markovianity becomes non-negligible~\cite{PhysRevA.86.012115}, which means the results from the Born-Marovian master equation method are unreliable. Our result is consistent with that of Ref.~\cite{doi:10.1080/00268976.2018.1430385} in which only the $z$-type coupling term is taken into consideration.

\section{Appendix B: Variational polaron transformation}

To obtain the explicit expression of $\tilde{\Omega}_{\mathrm{P}}$, we first need a rotation matrix $\mathcal{\hat{R}}\equiv\exp(-\frac{i}{2}\phi\hat{\sigma}_{y})$ with $\phi=\arctan(\cot\theta)$ to diagonalize $\mathcal{\hat{S}}$. By doing so, the original Hamiltonian $\hat{H}$ can be transformed into $\hat{H}'=\mathcal{\hat{R}}^{\dagger}\hat{H}\mathcal{\hat{R}}$, where
\setcounter{equation}{0}
\renewcommand\theequation{B\arabic{equation}}
\begin{equation}
\hat{H}'=\frac{1}{2}\tilde{\epsilon}\hat{\sigma}_{z}+\frac{1}{2} \tilde{\Delta}\hat{\sigma}_{x}+\hat{H}_{\mathrm{b}}+\hat{\sigma}_{z}\otimes\sum_{k}\tilde{g}_{k}(\hat{b}_{k}^{\dagger}+\hat{b}_{k}),
\end{equation}
with $\tilde{g}_{k}=\sin(\theta+\phi)g_{k}$. For the special scope $0\leq\theta<\pi$ considered in this paper, $\tilde{g}_{k}$ coincidentally reduces to $g_{k}$. Employing such rotation, $\hat{H}'$ now has the standard form of well-known spin-boson model.

Next, we apply the variational polaron transformation to $\hat{H}'$ as $\hat{H}''=\exp(\hat{\mathcal{P}})\hat{H}'\exp(-\hat{\mathcal{P}})$. Here, the polaron generator $\hat{\mathcal{P}}$ is given by~\cite{doi:10.1063/1.4722336}
\begin{equation}
\hat{\mathcal{P}}=\hat{\sigma}_{z}\sum_{k}\frac{\xi_{k}}{\omega_{k}}(\hat{b}_{k}^{\dagger}-\hat{b}_{k}),
\end{equation}
where $\xi_{k}=g_{k}\mathfrak{S}(\omega_{k})$ with
\begin{equation}
\mathfrak{S}(\omega_k)=\bigg{[}1+\frac{\eta^{2}\tilde{\Delta}^{2}}{\omega_{k}\tilde{\Omega}_{\mathrm{P}}}\coth\Big{(}\frac{1}{2}\beta\omega_{k}\Big{)}\tanh\Big{(}\frac{1}{2}\beta\tilde{\Omega}_{\mathrm{P}}\Big{)}\bigg{]}^{-1},
\end{equation}
being self-consistently determined by minimizing the Bogoliubov upper bound for the free energy~\cite{doi:10.1063/1.4722336}. The expression of $\hat{H}''$ can be written as $\hat{H}''=\hat{H}_{\mathrm{s}}''+\hat{H}_{\mathrm{b}}+\hat{H}_{\mathrm{sb}}''$, where $\hat{H}_{\mathrm{s}}''$ denotes the transformed Hamiltonian of the sensor
\begin{equation}
\hat{H}_{\mathrm{s}}''=\frac{1}{2}\tilde{\epsilon}\hat{\sigma}_{z}+\frac{1}{2} \eta\tilde{\Delta}\hat{\sigma}_{x}+\sum_{k}\frac{\xi_{k}}{\omega_{k}}\big{(}\xi_{k}-2g_{k}\big{)}.
\end{equation}
The last term in the above expression is an energy shift induced by the polaron transformation, which has no influence on the sensing accuracy in the assumption of canonical statistics. From the expression of $\hat{H}_{\mathrm{s}}''$, one can immediately find the Rabi frequency is $\tilde{\Omega}_{\mathrm{P}}^{2}=\tilde{\epsilon}^{2}+\eta^{2}\tilde{\Delta}^{2}$, which recovers Eq.~(\ref{eq:eq12}) in the main text. The transformed sensor-reservoir interaction Hamiltonian $\hat{H}_{\mathrm{sb}}''$ can be written as $\hat{H}_{\mathrm{sb}}''=\hat{\sigma}_{x}\mathcal{\hat{B}}_{x}+\hat{\sigma}_{y}\mathcal{\hat{B}}_{y}+\hat{\sigma}_{z}\mathcal{\hat{B}}_{z}$, where
\begin{equation}
\mathcal{\hat{B}}_{x}=\frac{1}{2}\tilde{\Delta}\big{(}\cosh\hat{\Sigma}-\eta\big{)},~~\mathcal{\hat{B}}_{y}=\frac{i}{2}\tilde{\Delta}\sinh\hat{\Sigma},
\end{equation}
\begin{equation}
\begin{split}
\mathcal{\hat{B}}_{z}=\sum_{k}\big{(}g_{k}-\xi_{k}\big{)}\Big{(}\hat{b}_{k}^{\dagger}+\hat{b}_{k}\Big{)},
\end{split}
\end{equation}
with
\begin{equation}
\hat{\Sigma}\equiv\sum_{k}\frac{2\xi_{k}}{\omega_{k}}\Big{(}\hat{b}_{k}^{\dagger}-\hat{b}_{k}\Big{)},	
\end{equation}
and
\begin{equation}\label{eq:eqb8}
\begin{split}
\eta\equiv&\mathrm{Tr}\Big{(}\varrho_{\mathrm{b}}^{\mathrm{G}}\cosh\hat{\Sigma}\Big{)}\\
=&\exp\bigg{[}-2\int_{0}^{\infty}d\omega\frac{J(\omega)}{\omega^{2}}\mathfrak{S}(\omega)^{2}\coth\Big{(}\frac{1}{2}\beta\omega\Big{)}\bigg{]}.
\end{split}
\end{equation}
\bibliography{reference}

\end{document}